\documentclass[prl,aps,twocolumn,amsmath,amssymb,superscriptaddress,showpacs]{revtex4}

\usepackage{graphicx}
\usepackage{bm}
\usepackage{upgreek}

\begin{document}

\title{Anomalous light cones and valley optical selection rules of interlayer excitons in twisted heterobilayers}

\author{Hongyi Yu}
\author{Yong Wang}
\author{Qingjun Tong}
\affiliation{Department of Physics and Center of Theoretical and Computational Physics, The University of Hong Kong, Hong Kong, China}

\author{Xiaodong Xu}
\affiliation{Department of Physics, University of Washington, Seattle, Washington 98195, USA}
\affiliation{Department of Materials Science and Engineering, University of Washington, Seattle, Washington 98195, USA}

\author{Wang Yao} \thanks{wangyao@hku.hk}
\affiliation{Department of Physics and Center of Theoretical and Computational Physics, The University of Hong Kong, Hong Kong, China}

\date{\today}

\begin{abstract}
We show that, because of the inevitable twist and lattice mismatch in heterobilayers of transition metal dichalcogenides, interlayer excitons have six-fold degenerate light cones anomalously located at finite velocities on the parabolic energy dispersion. The photon emissions at each light cone are elliptically polarized, with major axis locked to the direction of exciton velocity, and helicity specified by the valley indices of the electron and the hole. These finite-velocity light cones allow unprecedented possibilities to optically inject valley polarization and valley current, and the observation of both direct and inverse valley Hall effects, by exciting interlayer excitons. Our findings suggest potential excitonic circuits with valley functionalities, and unique opportunities to study exciton dynamics and condensation phenomena in semiconducting 2D heterostructures.
\end{abstract}

\pacs{74.78.Fk, 78.67.Pt, 71.35.-y, 72.25.Fe}
% 74.78.Fk	Multilayers, superlattices, heterostructures (under 74.78.-w	Superconducting films and low-dimensional structures)
% 78.67.Pt: Multilayers; superlattices; photonic structures; metamaterials (under 78.67.-n: Optical properties of low-dimensional, mesoscopic, and nanoscale materials and structures)
% 71.35.-y: Excitons and related phenomena
% 72.25.Fe: Optical creation of spin polarized carriers (under 72.25.-b: Spin polarized transport)

\maketitle

Monolayers of group-VIB transition metal dichalcogenides (TMDs) have recently emerged as a new class of direct-gap semiconductors in the two-dimensional (2D) limit \cite{Wang_NatNano,XuYao_Rev,GBLiu_Rev,Mak_PRL,Splendiani_NL}. These hexagonal 2D crystals have exotic properties associated with the valley degeneracy of the band edges, including the valley Hall effect \cite{Valley_SelectionRule,Mak_Vcurrent}, the valley magnetic moment \cite{Aivazian_MagnetoPL,Srivastava_MagnetoPL,Li_PRL,MacNeill_PRL}, and the valley optical selection rules \cite{Valley_SelectionRule,Yao_PRB,Mak_NatNano,Zeng_NatNano,Cao_NatCom,Jones_ValleyCoherence}, leading to rich possibilities for valley-based device applications. The visible range bandgap further makes these 2D semiconductors ideal platforms for optoelectronics. Due to the strong Coulomb interaction, the optical response is dominated by excitons, the hydrogen-like bound state of an electron-hole pair. The demonstrated electrostatic tunability and optical controllability of valley configurations of excitons in monolayer TMDs have implied new optoelectronic device concepts not possible in other material systems \cite{Mak_NatNano,Zeng_NatNano,Cao_NatCom,Jones_ValleyCoherence,Ross_NatCom}. 

Stacking different TMDs monolayers to form van der Waals heterostructures opens up a new realm to extend their already extraordinary properties \cite{Geim_Nature}. MoX$_2$/WX$_2$ (X=Se, S) heterobilayers have been realized \cite{Rivera_InterlayerX0,Lee_NatNano,Furchi_NL,Cheng_NL,Fang_PNAS,Hong_NatNano,Chiu_ACSNano}, which feature a type-II band alignment with the conduction (valence) band edges located in MoX$_2$ (WX$_2$) layer. Exciton then has the lowest energy in an interlayer configuration (i.e. electron and hole in different layers), from which luminescence is observed \cite{Rivera_InterlayerX0,Fang_PNAS,Chiu_ACSNano}. Due to the spatially indirect nature, interlayer excitons in MoSe$_2$/WSe$_2$ heterobilayers have shown long lifetime exceeding nanosecond, repulsive interaction, and electrostatically tunable resonance \cite{Rivera_InterlayerX0}, all of which are highly desirable for the realization of excitonic circuits and condensation \cite{Butov_Nature,Snoke_Nature,High_Science,Fogler_NatCom}. An unprecedented aspect of this interlayer exciton system is that the heterobilayers in general have incommensurate structures due to lattice mismatch and twist in the stacking which, together with the valley physics inherited from the monolayers, bring in radically new properties. 

Here we discover anomalous light coupling properties of interlayer excitons in twisted MoX$_2$/WX$_2$ heterobilayers. We find these excitons have unique six-fold degenerate light cones, located at finite center-of-mass velocities on the parabolic energy dispersion. In these light cones, an interlayer exciton can directly interconvert with an elliptically polarized photon (without phonon or impurity assistance), with helicity depending on the valley indices of the electron and the hole. These light cones allow resonant optical injection of excitonic valley polarization and valley current with versatile controllability, as well as the observation of valley Hall and inverse valley Hall effects of interlayer excitons. Contrary to known exciton systems, here the exciton lifetime is long at the energy minimum while optical injection and probe are allowed at finite velocities, suggesting unique opportunities to study exciton dynamics and condensation phenomena. For lattice matching heterobilayers of AA-like ($0^\circ$ twist) or AB-like ($60^\circ$ twist) stacking, all light cones merge into a single one, and the quantum interference leads to dependence of the brightness and the polarization selection rules on the translation between the layers.  

\begin{figure}
\includegraphics[width=\linewidth]{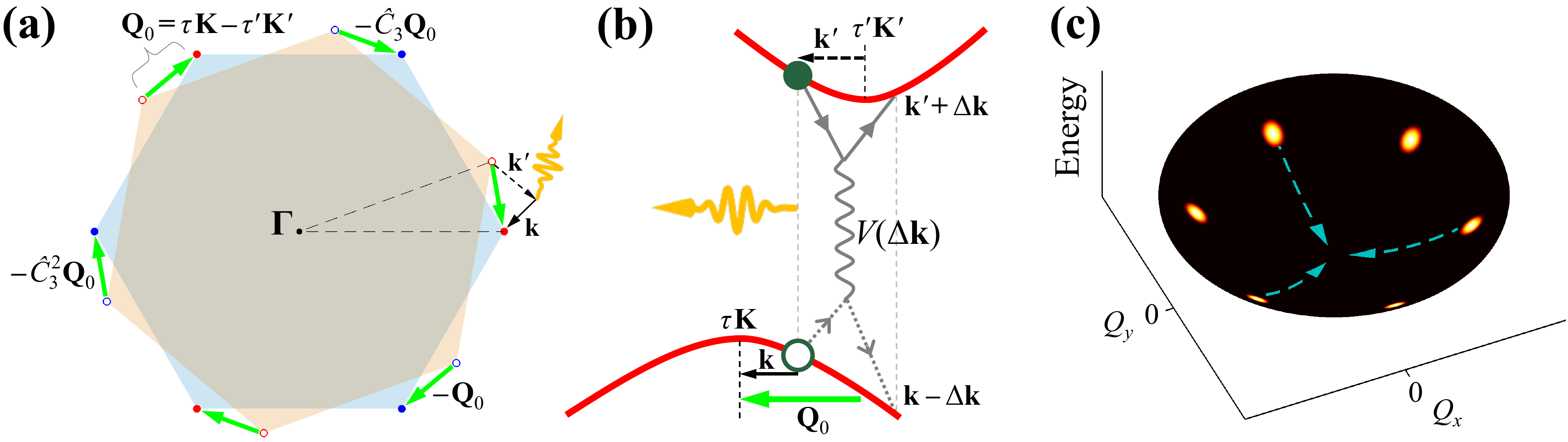}
\caption{(a) With a small twist and/or lattice mismatch, the Brillouin zone corners of the MoX$_2$ layer ($\tau'\mathbf K'$, open dots)  shift away from those of the WX$_2$ layer ($\tau\mathbf K$, solid dots). An electron and a hole can recombine if their momentum sum $\mathbf k + \mathbf k'$ equals the displacement vector between a $\tau\mathbf K$ and a $\tau'\mathbf K'$ corners (green arrows). (b) The interlayer Coulomb interaction (wavy line) between the electron and hole conserves their momentum sum ~\cite{Supplementary}. An interlayer exciton formed with kinematic momentum $\mathbf Q = \pm\mathbf Q_0$, $\pm \hat{C}_3 \mathbf Q_0$, or $\pm \hat{C}_3^2 \mathbf Q_0$ can directly recombine to emit photon. (c) These light cones at finite velocities are illustrated by the bright spots on the energy dispersion, while the exciton is optically dark at the energy minimum at $\mathbf Q = 0$. The dashed arrows illustrate the relaxation path ways.}
\label{fig1}
\end{figure}

\begin{figure}
\includegraphics[width=\linewidth]{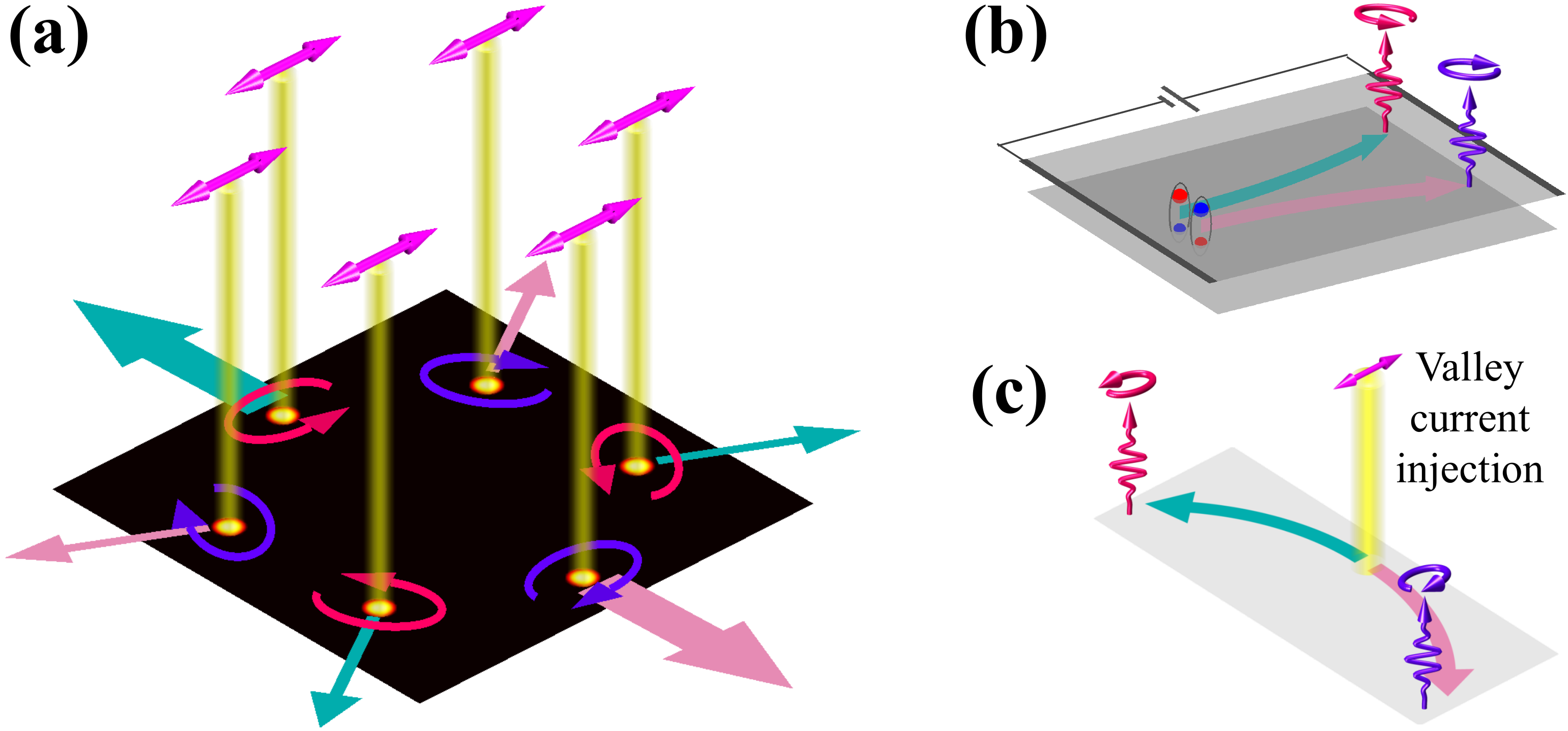}
\caption{(a) With the elliptically polarized selection rules, a linearly polarized light inject exciton currents with different rates and directions at the six light cones. The net effect is a pure valley current (green and pink colors denoting the valleys). (b) Valley Hall effect in heterobilayers. With the light cones at finite velocities, the spatial pattern of emission polarization also depends on the exciton distribution in $\mathbf Q$-space. (c) Exciton valley current can induce number current through the inverse valley Hall effect, observable from the spatial map of the luminescence.}
\label{fig2}
\end{figure}

The optically active interlayer excitons are the ones with electrons (holes) from the $\pm \mathbf K'$ ($\pm \mathbf K$) valleys in MoX$_2$ (WX$_2$), where layer-hybridization is substantially quenched by the large band offsets between the layers \cite{GBLiu_Rev,Chiu_HJBandAlignment,Debbichi_PRB,Kosmider_PRB,Lu_Nanoscale,Komsa_PRB,Kang_NL}. The term \emph{light cones} here refer to the exciton phase space regions where interconversion with photon is directly allowed (without phonon or impurity assistance). Concerning the light coupling, interlayer exciton in heterobilayers is distinct from existing systems by the fact that a small twist and/or lattice mismatch shifts the electron valleys away from the hole valleys (Fig. \ref{fig1}(a)). The exciton at zero velocity is therefore indirect in momentum space. The small mismatch between the electron and hole valleys can nevertheless be compensated by the exciton's kinematic momentum $\mathbf Q$ (Fig. \ref{fig1}(b)), to allow radiative recombination with a direct transition. This happens at $\mathbf Q= \pm \mathbf Q_0$, $\pm \hat{C}_3 \mathbf Q_0$, or $\pm \hat{C}_3^2 \mathbf Q_0$, where $\mathbf Q_0 \equiv \tau \mathbf K - \tau' \mathbf K '$ is the displacement from the MoX$_2$ Brillouin zone (BZ) corner $\tau' \mathbf K'$ to a nearest WX$_2$ BZ corner $\tau \mathbf K$, with $\tau'=\tau$ ($\tau'=-\tau$) if the twist angle $\theta$ is near $0^\circ$ ($60^\circ$). Thus, in the phase space parametrised by $\mathbf Q$, the light cones are all located at finite velocities with the six-fold degeneracy (Fig. \ref{fig1}(c)). With low symmetry of the exciton states, photon emission at each light cone can have a general form of being elliptically polarized (circular or linear polarization being the two limits when the ellipticity $\epsilon$ approaches $0$ or $1$). The exciton states at $\mathbf Q_0$, $\hat{C}_3 \mathbf Q_0$, and $ \hat{C}_3^2 \mathbf Q_0$ have the same electron-hole valley configuration $(\tau', \tau)$ and are related by the $\hat{C}_3$ rotation, so their elliptical polarizations are also $\hat{C}_3$ rotations of each other. The rest three light cones are their time-reversal counterparts with the valley configuration $(-\tau', -\tau)$ (Fig. \ref{fig2}(a)).

\textit{Optical injection of valley current.}---With the helicity determined by the valley index and the major axis of the elliptical polarization locked to the exciton moving direction, the finite-velocity light cones make possible valley selective injection of valley polarization as well as valley current (Fig. \ref{fig2}(a)). Under excitation by a linearly polarized light, the absorption rates are all different at the six light cones associated with different velocities, giving rise to valley current (Fig. \ref{fig2}(a)). Using $\phi$ to denote the angle between the major axis of polarization of the light cone at $\mathbf Q_0=Q_0\mathbf y$ and that of a linearly polarized pumping light, a pure valley current is injected with rate $\frac{3}{4}\frac{1-(1-\epsilon)^2}{1+(1-\epsilon)^2}\frac{\hbar K\theta}{M_0}\alpha$ in the direction $\mathbf x\sin 2\phi+\mathbf y\cos 2\phi$, $\alpha$ being the population injection rate. 

\textit{Optical valley Hall effect.}---This exciton system is an ideal platform to study the Berry phase effect in the Bloch band. TMD monolayers have valley dependent Berry curvatures in their conduction and valence band edges, which give rise to the valley Hall effect of the carriers \cite{Valley_SelectionRule,Mak_Vcurrent}. The exciton can inherit the Berry curvature of its electron and hole constituents \cite{Yao_X0Transport,XCurvature},
giving rise to an exciton valley Hall effect that is observable from the spatial pattern of emission polarization (Fig. \ref{fig2}(b)) \cite{Yao_X0Transport,Leyder_NatPhys}. The elliptically polarized emission on the two edges with opposite helicity is a phenomenon unique to this heterobilayer interlayer exciton system. The optical injection of excitonic valley current through the finite-velocity light cones further makes possible the observation of inverse valley Hall effect, where the valley current induces a number current in perpendicular direction due to the skewed motion by the Berry curvature (Fig. \ref{fig2}(c)). 

\begin{figure*}
\includegraphics[width=\linewidth]{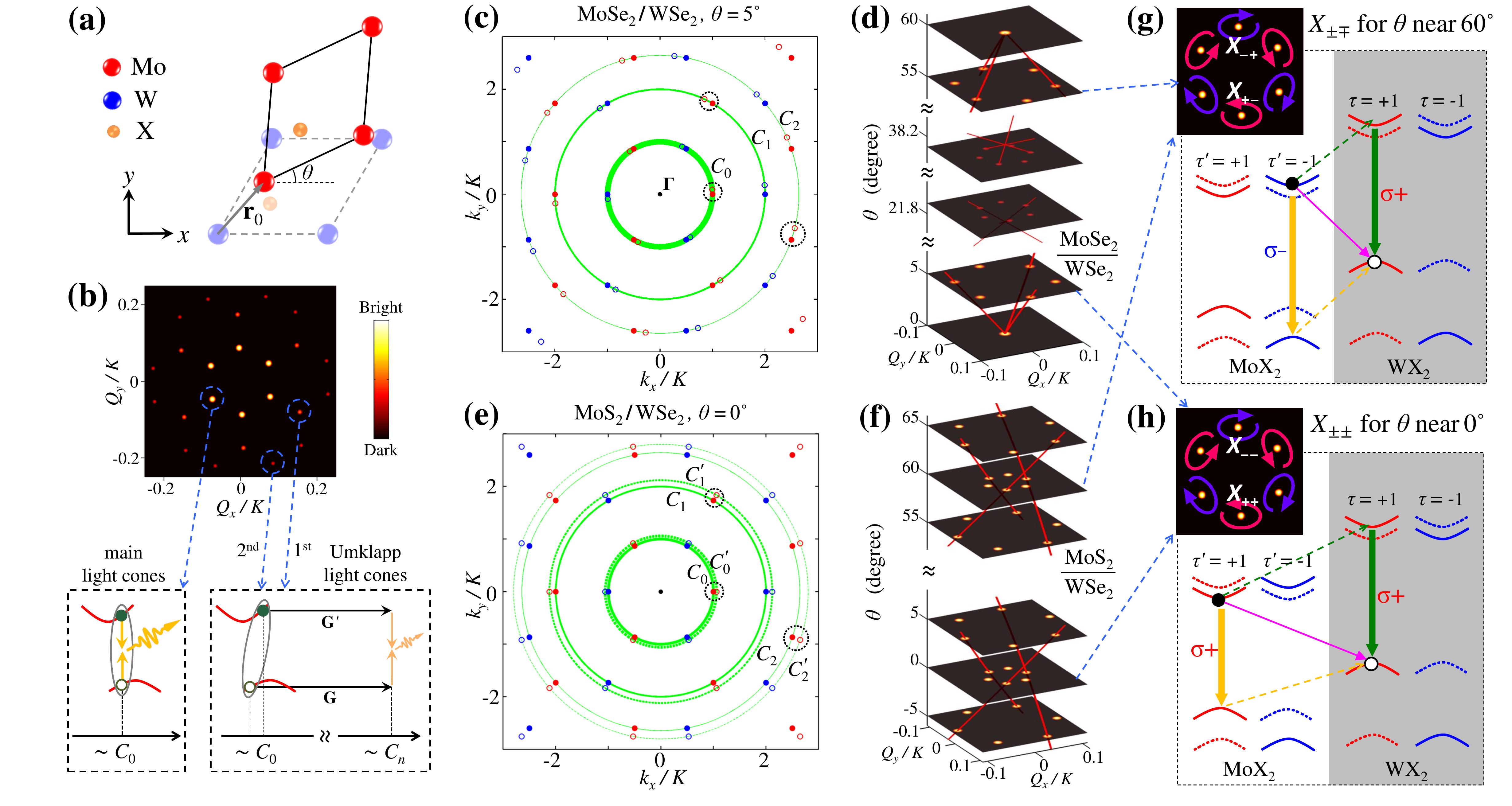}
\caption{(a) A MoX$_2$/WX$_2$ heterobilayer can be specified by the twist angle $\theta$ and the interlayer translation $\mathbf r_0$. (b) The bright spots from large to small denote respectively the main, 1st Umklapp, and 2nd Umklapp light cones in the $\mathbf Q$-space, for WSe$_2$/MoSe$_2$ heterobilayers with $\theta$ near $0^\circ$ or $60^\circ$. (c) Solid (open) dots denote $\tau\mathbf K+\mathbf G$ ($\tau'\mathbf K'+\mathbf G'$) points of WSe$_2$ (MoSe$_2$) layer in the extended zone scheme for $\theta = 5^\circ$. The locations of main ($n$th Umklapp) light cones are given by the displacements between nearest solid and open dots on the $C_0$ ($C_n$) circles. (d) Low energy light cones at $\theta=0^\circ, 5^\circ, 22.8^\circ, 37.2^\circ, 55^\circ, 60^\circ$, respectively. Evolutions of the main (2nd Umklapp) light cones with $\theta$ are shown by the thick (thin) red lines.
(e), (f) Lattice mismatched MoS$_2$/WSe$_2$ heterobilayer. All light cones are always at finite $\mathbf Q$. (g) Recombination pathways of interlayer exciton $X_{-+}$ at $\theta$ near $60^\circ$. Solid (dashed) arrows denote dipole transition (interlayer hopping). The inset shows the elliptical polarizations of emissions at the main light cones. (h) $X_{++}$ at $\theta$ near $0^\circ$. At the finite $\mathbf Q$ light cones, the polarization and strength of the optical dipole are independent of $\mathbf r_0$.}
\label{fig3}
\end{figure*}

We present now the theoretical details that backup the above general statements on the anomalous light cones. The heterobilayers can be characterized by the twist angle $0^{\circ}\le\theta\le60^{\circ}$ and the in-plane translation $\mathbf r_0$ between the layers (Fig. \ref{fig3}(a)). The origin of the in-plane coordinates is chosen at a Mo atom, and the coordinate of its nearest W atom defines $\mathbf r_0$ \cite{Supplementary}. 
We first give the description of interlayer valley excitons in twisted heterobilayers at vanishing interlayer coupling, and then analyze their light-interaction effects due to the residue interlayer coupling. 
Using $|e_{\tau',\mathbf k'}\rangle$ ($|h_{\tau,\mathbf k}\rangle$) to denote the electron (hole) Bloch state at momentum $\mathbf k'$ ($\mathbf k$) measured from $\tau'\mathbf K'$ ($\tau\mathbf K$) in MoX$_2$ (WX$_2$) layer in the \textit{absence} of interlayer hybridization, the exciton wavefunction writes \cite{Snoke_X0BEC,Supplementary}
\begin{equation}
|X^{(0)}_{\tau'\tau,\mathbf Q}\rangle=\sum_{\Delta\mathbf Q}\Phi_I (\Delta\mathbf Q)|e_{\tau',\frac{m_e\mathbf Q}{M_0}+\Delta\mathbf Q}\rangle|h_{\tau,\frac{m_h\mathbf Q}{M_0}-\Delta\mathbf Q}\rangle,
\label{InterlayerX0WF}
\end{equation}
with the eigenenergy $E_I(\mathbf Q)=\frac{\hbar^2Q^2}{2M_0}+\Delta_g-E_b$. $M_0\equiv m_e+m_h$ is the exciton mass, $\Delta_g$ the band-gap, $E_b$ the binding energy, and $\Phi_I(\Delta\mathbf Q)$ describes the electron-hole relative motion with $\Delta\mathbf Q\equiv\frac{m_h\mathbf k'-m_e\mathbf k}{M_0}$. Besides the electron and hole valley indices $\tau'$ and $\tau$, an interlayer exciton state is labeled by $\mathbf Q\equiv\mathbf k+\mathbf k'$, the kinematical momentum associated with the center-of-mass group velocity $\langle X^{(0)}_{\mathbf Q}|(\frac{m_e}{M_0}\dot{\mathbf r}_e+\frac{m_h}{M_0}\dot{\mathbf r}_h)|X^{(0)}_{\mathbf Q}\rangle=\frac{\hbar}{M_0}\mathbf Q$~\cite{Supplementary}, which is a good quantum number even in absence of translational invariance at general twist angles. 

The Fourier expansion of the electron (hole) Bloch function consists of the plane-wave components $e^{i(\mathbf k'+\tau'\mathbf K'+\mathbf G')\cdot\mathbf r}$ ($e^{i(\mathbf k-\tau\mathbf K-\mathbf G)\cdot\mathbf r}$), where $\mathbf G'$ ($\mathbf G$) denotes all possible reciprocal lattice vectors in the MoX$_2$ (WX$_2$) layer. Neglecting the small photon wave vector, an electron-hole pair $|e_{\tau',\mathbf k'}\rangle|h_{\tau,\mathbf k}\rangle$ can recombine only when $\mathbf k'+\tau'\mathbf K'+\mathbf G' = - \mathbf k + \tau\mathbf K + \mathbf G$. So the exciton light cones are at the $\mathbf Q$-space points:
\begin{equation}
\mathbf Q=\tau\bm\upkappa-\tau'\bm\upkappa',
\label{LightCone}
\end{equation}
where $\tau'\bm\upkappa'\equiv\tau'\mathbf K'+\mathbf G'$ ($\tau\bm\upkappa\equiv\tau\mathbf K+\mathbf G$).

In Fig. \ref{fig3}(c) and (e), $\tau\bm\upkappa$ and $\tau'\bm\upkappa'$ are shown in the extended zone scheme for a MoSe$_2$/WSe$_2$ heterobilayer at $\theta=5^\circ$ and a MoS$_2$/WSe$_2$ heterobilayer at $\theta=0^\circ$. When $\theta$ changes, $\tau'\bm\upkappa'$ ($\tau\bm\upkappa$) move on concentric circles denoted as $C_n'$ ($C_n$). The nearest $(\tau\bm\upkappa,\tau'\bm\upkappa')$ pairs on $C_0$ and $C_0'$ correspond to main light cones. And the nearest $(\tau\bm\upkappa,\tau'\bm\upkappa')$ pairs on $C_n$ and $C_n'$ define the $n$th \emph{Umklapp light cones}, where light coupling is an Umklapp type process (Fig. \ref{fig3}(b)). Fig. \ref{fig3}(d) and (f) shows the locations of these light cones as a function of twist angle $\theta$, in a range $Q\le0.1K$ that corresponds to a kinetic energy $\frac{\hbar^2Q^2}{2M_0}\lesssim60$ meV. The six main light cones are located at $\mathbf Q_0=(K-K'\cos\delta\theta)\mathbf x-K'\sin\delta\theta\mathbf y$ and its $\frac{n\pi}{3}$ rotations, where $\delta\theta$ is the deviation of $\theta$ from $0^\circ$ or $60^\circ$. Three of them correspond to the same valley configurations (red lines in Fig. \ref{fig3}(d) and (f)), and their time-reversals give the remaining three. 1st-Umklapp light cones also have six-fold degeneracy, located at $2\mathbf Q_0$. The main and 1st-Umklapp light cones are the lowest energy ones at $\theta$ near $0^\circ$ or $60^\circ$. There are twelve 2nd-Umklapp light cones (Fig. \ref{fig3}(b)), and six of them become the lowest energy ones for lattice matching heterobilayers when $\theta$ is near the commensurate angle $21.8^\circ$ or $38.2^\circ$ (Fig. \ref{fig3}(d)). 

At these light cones, the exciton acquires a finite optical transition dipole from the interlayer coupling. The interlayer hoppings ($\hat H_T$) of electron and hole will correct the interlayer exciton wavefunction in Eq. (\ref{InterlayerX0WF}), hybridizing $| X^{(0)}_{\tau'\tau,\mathbf Q}\rangle$ with small components of the \emph{intralayer} valley excitons~\cite{Supplementary}. The optical transition matrix element of the interlayer exciton is then:
\begin{align}
\bm{\mathcal{D}}_{\tau'\tau,\mathbf Q}=&\langle0|\hat{\mathbf D}|X^{(0)}_{\tau'\tau,\mathbf Q}\rangle+ \langle0|\hat{\mathbf D}|X_{\tau'}\rangle \frac{\langle X_{\tau'}|\hat H_T | X^{(0)}_{\tau'\tau,\mathbf Q}\rangle}{E_I(\mathbf Q)-E_M} \nonumber\\
&+\langle0|\hat{\mathbf D}|X_{\tau}\rangle \frac{\langle X_\tau|\hat H_T | X^{(0)}_{\tau'\tau,\mathbf Q}\rangle}{E_I(\mathbf Q)-E_W}.
\label{OpticalDipole}
\end{align}
Here $\hat{\mathbf D}$ is the electric dipole operator and $|0\rangle$ is the vacuum state. The first term on the right-hand-side is from the transition dipole between $|e_{\tau',\mathbf k'}\rangle$ and $|h_{\tau,\mathbf k}\rangle$ which is small because of the spatial separation. The second and third terms describe respectively the light coupling mediated by bright intralayer exciton $X_{\tau'}$ in MoX$_2$ layer through hole interlayer hopping, and $X_{\tau}$ in WX$_2$ through electron hopping. $E_{M}$ and $E_{W}$ are the energies of the intralayer excitons, a few hundred meV above $E_{I}$ as determined from optical and scanning tunneling measurements~\cite{Rivera_InterlayerX0,Fang_PNAS,Chiu_ACSNano,Chiu_HJBandAlignment}. 
$\bm{\mathcal{D}}_{\tau'\tau,\mathbf Q}$ can be extracted from \textit{ab initio} calculations at several commensurate stackings~\cite{Supplementary}, from which we can determine the dipole strength, and the ellipticity and major axes of the emission polarization of the six-fold light cones at a general twist angle.

In the two-center approximation~\cite{TwistedGrapheneTheo,TwistedGrapheneExp}:
\begin{align}
\langle0|\hat{\mathbf D}|X^{(0)}_{\tau'\tau,\mathbf Q}\rangle &\propto \sum_{\bm\upkappa,\bm\upkappa'}\delta_{\mathbf Q,\tau\bm\upkappa-\tau'\bm\upkappa'}\mathbf D_0^{2\tau}(\tau\bm\upkappa)e^{-i\tau\bm\upkappa\cdot\mathbf r_0},\nonumber\\
\langle X_{\tau'}|\hat H_T|X^{(0)}_{\tau'\tau,\mathbf Q}\rangle &\approx \sum_{\bm\upkappa,\bm\upkappa'}\delta_{\mathbf Q,\tau\bm\upkappa-\tau'\bm\upkappa'}t_{2\tau'}^{2\tau}(\tau\bm\upkappa)e^{-i\tau\bm\upkappa\cdot\mathbf r_0},\nonumber\\
\langle X_\tau|\hat H_T|X^{(0)}_{\tau'\tau,\mathbf Q}\rangle  &\approx \sum_{\bm\upkappa,\bm\upkappa'}\delta_{\mathbf Q,\tau\bm\upkappa-\tau'\bm\upkappa'}t_0^0(\tau\bm\upkappa)e^{-i\tau\bm\upkappa\cdot\mathbf r_0},
\label{IntraInterCoupling}
\end{align}
where $t_{m'}^m$ ($\mathbf D_{m'}^m$) are the Fourier components of hopping integral (transition dipole) between a W and a Mo $d$-orbitals with magnetic quantum numbers $m$ and $m'$ respectively, which decay fast with the increase of $|\bm\upkappa|$~\cite{Supplementary}. 

Fig. \ref{fig3}(h) schematically shows interlayer excitons $X_{\tau'\tau}$ with valley configuration $\tau'=\tau$. These are the ones in the main and 1st Umklapp light cones at $\theta$ near $0^\circ$ and in 2nd Umklapp light cones at $\theta$ near $38.2^\circ$. Fig. \ref{fig3}(g) shows $X_{\tau'\tau}$ with $\tau'=-\tau$, which are the ones in the main and 1st Umklapp light cones at $\theta$ near $60^\circ$ and in 2nd Umklapp light cones at $\theta$ near $21.8^\circ$. The emission polarizations of the main light cones are illustrated in the insets. For MoSe$_2$/WSe$_2$ heterobilayers, we find the emission polarization of $X_{++}$ at $\theta$ near $0^\circ$ has an ellipticity $\epsilon=0.35$, which correspond to a ratio of $5$ between the $\mathbf{e}_+ \equiv \frac{\mathbf x + i\mathbf y}{\sqrt{2}}$ and $\mathbf{e}_- \equiv \frac{ \mathbf x - i\mathbf y}{\sqrt{2}}$ component of $\bm{\mathcal{D}}_{++,\mathbf Q}$. The $\mathbf{e}_+$ component is contributed by all three terms in Eq. (\ref{OpticalDipole}), while the much smaller $\mathbf{e}_-$ component is from the $\langle0|\hat{\mathbf D}|X^{(0)}_{\tau'\tau,\mathbf Q}\rangle$ term only. $X_{--}$ is simply the time reversal of $X_{++}$. At $\theta$ near $60^\circ$, the valley configurations $X_{+-}$ and $X_{-+}$ have the ellipticity $\epsilon = 0.33$. 

From \textit{ab initio} calculations \cite{Supplementary}, $X_{\pm \pm}$ at $\theta$ near $0^\circ$ has a transition dipole strength of $\sim 0.052D$ at the main light cones, and $\sim 0.001D$ at both the 1st and 2nd Umklapp light cones. $X_{\pm \mp}$ at $\theta$ near $60^\circ$ has a transition dipole strength of $\sim 0.052D$ at the main light cones, $\sim 0.001D$ at 1st Umklapp, and $\sim 0.0005D$ at 2nd Umklapp light cones. $D$ is the dipole strength of the intralayer excitons \cite{Supplementary,MoSe2RabiSplitting}. Eq. (\ref{IntraInterCoupling}) shows that the strength and polarization of $\bm{\mathcal{D}}_{\tau'\tau}$ at the six-fold light cones have no dependence on the layer translation $\mathbf r_0$ which appears in a phase factor $e^{-i\tau\bm\upkappa\cdot\mathbf r_0}$ only.

\begin{figure}
\includegraphics[width=\linewidth]{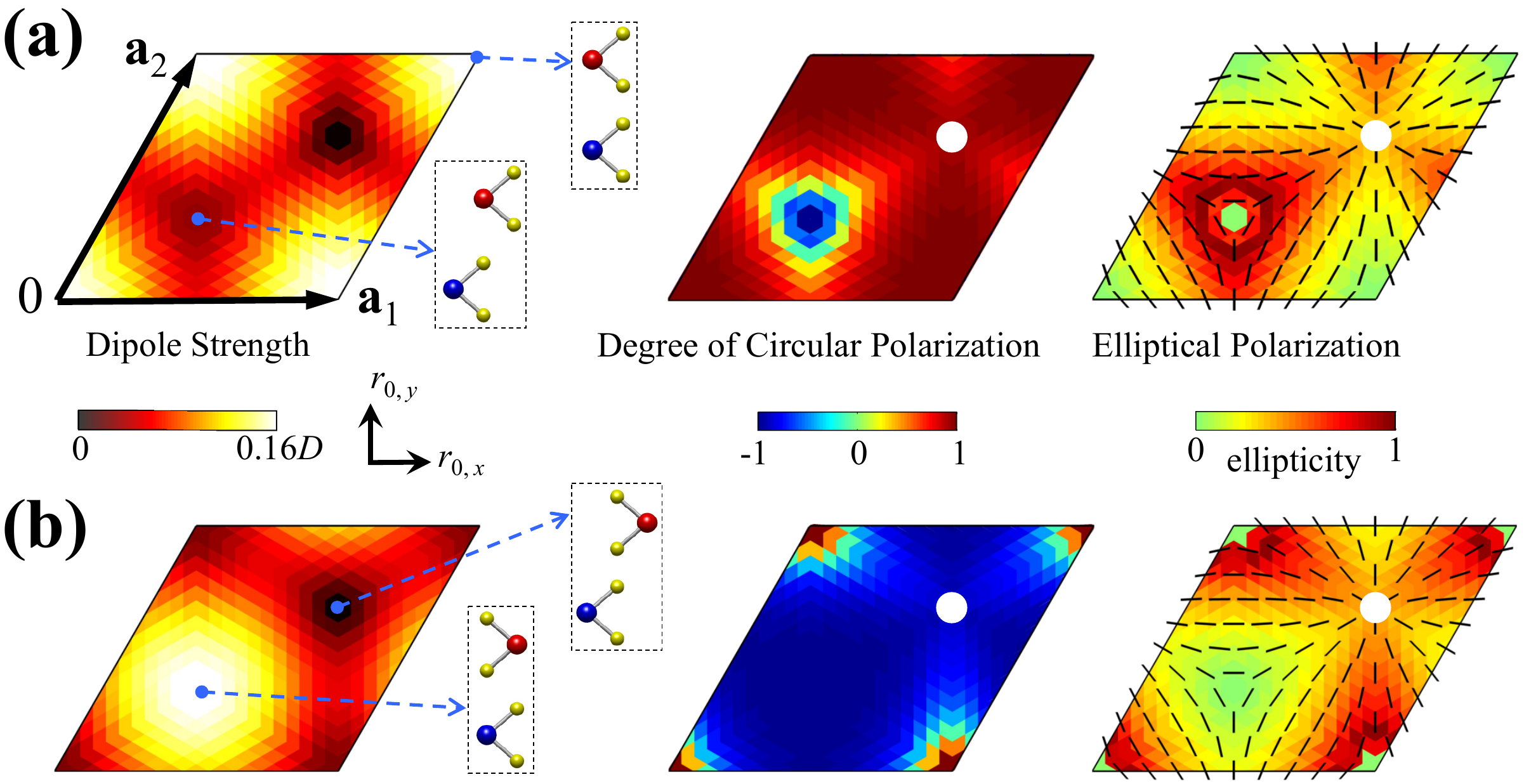}
\caption{Dependence of the transition dipole strength and polarization as a function of layer translation $\mathbf r_0$ in lattice matching MoSe$_2$/WSe$_2$ heterobilayers. (a)  $X_{++}$ at AA-like stacking ($\theta=0^\circ$). The three panels show respectively the dipole strength, the degree of polarization in circular basis, and the ellipticity and major axis (representd by short lines) of elliptical polarization, from {\it ab initio} calculations~\cite{abinitio}. (b) Same plots for $X_{-+}$ at AB-like stacking ($\theta=60^\circ$).}
\label{fig4}
\end{figure}

\textit{Translation dependence at commensurate stacking.}---For MoSe$_2$/WSe$_2$ or MoS$_2$/WS$_2$ heterobilayers, neglecting the $\sim0.1\%$ lattice mismatch, they have commensurate stacking at $\theta=0^\circ$, $60^\circ$, $21.8^\circ$ or $38.2^\circ$. At $\theta=0^\circ, 60^\circ$, all main and Umklapp light cones merge into a single cone at $\mathbf Q=0$. At $\theta=21.8^\circ, 38.2^\circ$, the six 2nd Umklapp light cones merge into a single one at $\mathbf Q=0$ (c.f. Fig. \ref{fig3}(d)). For the light cone at $\mathbf Q=0$, the optical transition dipole is the superposition of contributions from multiple $(\tau\bm\upkappa,\tau'\bm\upkappa')$ pairs each of which is associated with a distinct phase factor $e^{-i\tau\bm\upkappa\cdot\mathbf r_0}$ (c.f. Eq. (\ref{IntraInterCoupling})). Their interference then gives rise to the different $\mathbf r_0$ dependence for the $\sigma+$ and $\sigma-$ components of $\bm{\mathcal{D}}_{\tau'\tau}$~\cite{Supplementary}: 
\begin{align}
\mathbf{e}_\tau\!\cdot\!\bm{\mathcal{D}}_{\tau'\tau}&\propto e^{-i\tau\mathbf K\cdot\mathbf r_0}+e^{-i \tau\hat C_{3}\mathbf K\cdot\mathbf r_0}+e^{-i\tau\hat C_{3}^2\mathbf K\cdot\mathbf r_0}, \\
\mathbf{e}_{-\tau}\!\cdot\!\bm{\mathcal{D}}_{\tau'\tau}&\propto e^{-i \tau\mathbf K\cdot\mathbf r_0} + e^{-i \tau(\hat C_{3}\mathbf K\cdot\mathbf r_0 + \frac{2\pi}{3})}+ e^{-i \tau(\hat C_{3}^2\mathbf K\cdot\mathbf r_0+\frac{4\pi}{3})} \notag
\end{align}
Thus both the strength and the elliptical polarization of the transition dipole change with $\mathbf r_0$~\cite{layerseparation}, as shown in Fig. \ref{fig4}. The above relations dictate that at $\mathbf r_0=0$ and $\frac{\mathbf a_1+\mathbf a_2}{3}$ the in-plane transition dipole of $X_{\tau'\tau}$ becomes fully circularly polarized, while at $\mathbf r_0=\frac{2(\mathbf a_1+\mathbf a_2)}{3}$ both $\mathbf{e}_{\pm}$ components of $\bm{\mathcal{D}}_{\tau'\tau}$ vanish, in agreement with the \textit{ab initio} calculations in Fig.~\ref{fig4}~\cite{abinitio}. Only at these three $\mathbf r_0$, the commensurate heterobilayers have the $3$-fold rotational symmetry. For other $\mathbf r_0$, this rotational symmetry is absent and elliptically polarized emission by $X_{\tau'\tau}$ at $\mathbf Q=0$ is allowed. For $\theta=0^\circ$, the transition dipole is found to be strongest $\sim 0.16D$ at $\mathbf r_0=0$ (i.e. AA stacking, c.f. Fig. \ref{fig4}(a)), while for $\theta=60^\circ$, the transition dipole is strongest $\sim 0.15D$ at $\mathbf r_0=\frac{\mathbf a_1+\mathbf a_2}{3}$ (i.e. AB stacking, c.f. Fig. \ref{fig4}(b)).

\textit{Discussions.}---For interlayer excitons relaxed to the bottom of the energy dispersion, their photon emission need the assistance of phonon or impurity scattering into the light cones at the finite velocities. The inefficiency of such phonon or impurity assisted radiative recombination process, compared to the direct recombination of intralayer excitons, can be relevant to the long lifetime (exceeding nanosecond) observed in the time resolved photoluminescence (PL) \cite{Rivera_InterlayerX0}, which can be interesting for future study. The observation at the same time implies the inefficiency of non-radiative recombination of interlayer excitons. The energy of the emitted photon is $E_I(\mathbf Q)$ plus (minus) the phonon energy if phonon absorption (emission) is involved. The distribution of exciton kinetic energy and the phonon energy can account for the large broadening of PL. Our results also explain why strong interlayer exciton PL is observed only at twist angle $\theta$ near $0^\circ$ or $60^\circ$ \cite{Rivera_InterlayerX0}, where the main light cones get sufficiently close to the bottom of energy dispersion. There are other twist angles ($\theta$ near $21.8^\circ$ or $38.2^\circ$) where the 2nd Umklapp light cones are close to $\mathbf Q=0$, however, the weak dipole of these light cones renders the PL inefficient. 

\textit{Acknowledgments.}---This work is mainly supported by the Croucher Foundation (Croucher Innovation Award), the Research Grant Council (HKU17305914P) and University Grant Committee (AoE/P-04/08) of HKSAR, and University of Hong Kong (OYRA and ROP). X.X. is supported by NSF (DMR-1150719 and EFRI-1433496), and Cottrell Scholar Award.


\begin{thebibliography}{99}
\bibitem{Wang_NatNano}
Q. H. Wang \textit{et al.}, Nature Nanotech. $\bf 7$, 699 (2012).

\bibitem{XuYao_Rev}
X. Xu, W. Yao, D. Xiao, and T. F. Heinz, Nature Phys. $\bf 10$, 343 (2014).

\bibitem{GBLiu_Rev}
G.-B. Liu \textit{et al.}, Chem. Soc. Rev. $\bf 44$, 2643 (2015).

\bibitem{Mak_PRL}
K. F. Mak, C. Lee, J. Hone, J. Shan, and T. F. Heinz, Phys. Rev. Lett. $\bf 105$, 136805 (2010).

\bibitem{Splendiani_NL}
A. Splendiani \textit{et al.}, Nano Lett. $\bf 10$, 1271 (2010).

\bibitem{Valley_SelectionRule}
D. Xiao, G.-B. Liu, W. Feng, X. Xu, and W. Yao, Phys. Rev. Lett. $\bf 108$, 196802 (2012).

\bibitem{Mak_Vcurrent}
K. F. Mak, K. L. McGill, J. Park, and P. L. McEuen, Science $\bf 344$, 1489 (2014).

\bibitem{Aivazian_MagnetoPL}
G. Aivazian \textit{et al.}, Nature Phys. $\bf 11$, 148 (2015).

\bibitem{Srivastava_MagnetoPL}
A. Srivastava \textit{et al.}, Nature Phys. $\bf 11$, 141 (2015).

\bibitem{Li_PRL}
Y. Li \textit{et al.}, Phys. Rev. Lett. $\bf 113$, 266804 (2014).

\bibitem{MacNeill_PRL}
D. MacNeill \textit{et al.}, Phys. Rev. Lett. $\bf 114$, 037401 (2015).

\bibitem{Yao_PRB}
W. Yao, D. Xiao, and Q. Niu, Phys. Rev. B $\bf 77$, 235406 (2008).

\bibitem{Mak_NatNano}
K. F. Mak, K. He, J. Shan, and T. F. Heinz, Nature Nanotech. $\bf 7$, 494 (2012).

\bibitem{Zeng_NatNano}
H. Zeng \textit{et al.}, Nature Nanotech. $\bf 7$, 490 (2012).

\bibitem{Cao_NatCom}
T. Cao \textit{et al.}, Nat. Commun. $\bf 3$, 887 (2012).

\bibitem{Jones_ValleyCoherence}
A. M. Jones \textit{et al.}, Nature Nanotech. $\bf 8$, 634 (2013).

\bibitem{Ross_NatCom}
J. S. Ross \textit{et al.}, Nat. Commun. $\bf 4$, 1474 (2012).

\bibitem{Geim_Nature}
A. K. Geim and I. V. Grigorieva, Nature $\bf 499$, 419 (2013).

\bibitem{Rivera_InterlayerX0}
P. Rivera \textit{et al.}, Nat. Commun. $\bf 6$, 6242 (2015).

\bibitem{Fang_PNAS}
H. Fang \textit{et al.}, Proc. Natl. Acad. Sci. USA $\bf 111$, 6198 (2014).

\bibitem{Chiu_ACSNano}
M.-H. Chiu \textit{et al.}, ACS Nano $\bf 8$, 9649 (2014).

\bibitem{Lee_NatNano}
C.-H. Lee \textit{et al.}, Nature Nanotech. $\bf 9$, 676 (2014).

\bibitem{Furchi_NL}
M. M. Furchi \textit{et al.}, Nano Lett. $\bf 14$, 4785 (2014).

\bibitem{Cheng_NL}
R. Cheng \textit{et al.}, Nano Lett. $\bf 14$, 5590 (2014).

\bibitem{Hong_NatNano}
X. Hong \textit{et al.}, Nature Nanotech. $\bf 9$, 682 (2014).

\bibitem{Butov_Nature}
L. V. Butov, A. C. Gossard, and D. S. Chemla, Nature $\bf 418$, 751 (2002).

\bibitem{Snoke_Nature}
D. Snoke \textit{et al.}, Nature $\bf 418$, 754 (2002).

\bibitem{High_Science}
A. A. High \textit{et al.}, Science $\bf 321$, 229 (2008).

\bibitem{Fogler_NatCom}
M. M. Fogler, L. V. Butov, and K. S. Novoselov, Nat. Commun. $\bf 5$, 4555 (2014).
 
\bibitem{Chiu_HJBandAlignment}
M.-H. Chiu \textit{et al.}, Nat. Commun. $\bf 6$, 7666 (2015).

\bibitem{Debbichi_PRB}
L. Debbichi, O. Eriksson, and S. Leb{\'e}gue, Phys. Rev. B $\bf 89$, 205311 (2014).

\bibitem{Kosmider_PRB}
K. Ko{\'s}mider and J. Fern{\'a}ndez-Rossier, Phys. Rev. B $\bf 87$, 075451 (2013).

\bibitem{Lu_Nanoscale}
N. Lu \textit{et al.}, Nanoscale $\bf 6$, 2879 (2014).

\bibitem{Komsa_PRB}
H.-P. Komsa and A. V. Krasheninnikov, Phys. Rev. B $\bf 88$, 085318 (2013).

\bibitem{Kang_NL}
J. Kang \textit{et al.}, Nano Lett. $\bf 13$, 5485 (2013).

\bibitem{Yao_X0Transport}
W. Yao and Q. Niu, Phys. Rev. Lett. $\bf 101$, 106401 (2008).

\bibitem{XCurvature}
The Berry curvature of interlayer exciton $X_{\tau' \tau}$ is $\tau' \Omega_e + \tau \Omega_h$. The electron and hole Berry curvatures $\Omega_e$ and $\Omega_h$ from the MoX$_2$ and WX$_2$ respectively are both $\sim O(10)$ \AA$^2$, but can have a difference of $\sim50\%$ \cite{GBLiu_Rev}.

\bibitem{Leyder_NatPhys}
C. Leyder \textit{et al.}, Nature Phys. $\bf 3$, 628 (2007).

\bibitem{Snoke_X0BEC}
S. A. Moskalenko and D. W. Snoke, \textit{Bose-Einstein Condensation of Excitons and Biexcitons} (Cambridge University Press, 2000).

\bibitem{Supplementary}
See the Supplemental Material for the detailed analysis of interlayer Coulomb interaction, interlayer hopping and transition dipole amplitude and \textit{ab initio} calculations, which includes Refs. \cite{Yu_NatCom, Bassani_Book, Qiu_X0WF, Berkelbach_EffMass, Giannozzi_JPCM, Perdew_PRL, Bhattacharyya_PRB}.

\bibitem{Yu_NatCom}
H. Yu \textit{et al.}, Nat. Commun. $\bf 5$, 3876 (2014).

\bibitem{Bassani_Book}
F. Bassani and G. P. Parravicini, \textit{Electronic states and optical transitions in solids} (Pergamon Press, 1975).

\bibitem{Qiu_X0WF}
D. Y. Qiu, F. H. d. Jornada, and S. G. Louie, Phys. Rev. Lett. $\bf 111$, 216805 (2013).

\bibitem{Berkelbach_EffMass}
T. C. Berkelbach, M. S. Hybertsen, and D. R. Reichman, Phys. Rev. B $\bf 88$, 045318 (2013).

\bibitem{Giannozzi_JPCM}
P. Giannozzi \textit{et al.}, J. Phys. Condens. Matter $\bf 21$, 395502 (2009).

\bibitem{Perdew_PRL}
J. P. Perdew, K. Burke, and M. Ernzerhof, Phys. Rev. Lett. $\bf 77$, 3865 (1996).

\bibitem{Bhattacharyya_PRB}
S. Bhattacharyya and A. K. Singh, Phys. Rev. B $\bf 86$, 075454 (2012).

\bibitem{TwistedGrapheneTheo}
R. Bistritzer and A. H. MacDonald, Phys. Rev. B $\bf 81$, 245412 (2010).

\bibitem{TwistedGrapheneExp}
A. Mishchenko \textit{et al.}, Nature Nanotech. $\bf 9$, 808 (2014).

\bibitem{MoSe2RabiSplitting}
S. Dufferwiel et al., arXiv:1505.04438 (2015).

\bibitem{layerseparation}
The interlayer separation also varies with $\mathbf r_0$ in reality which can affect the dipole strength \cite{Zande_NL,Liu_NatCom}, but not the selection rule.

\bibitem{Zande_NL}
A. M. v. d. Zande \textit{et al.}, Nano Lett. $\bf 14$, 3869 (2014).

\bibitem{Liu_NatCom}
K. Liu \textit{et al.}, Nat. Commun. $\bf 5$, 4966 (2014).

\bibitem{abinitio}
In commensurate heterobilayers, the interlayer exciton transition dipole is given from the \textit{ab initio} calculations of the band-to-band dipole matrix elements at $\mathbf K$, normalized by the exciton envelope function. We have assumed the latter does not depend significantly on $\mathbf r_0$. 

\end{thebibliography}
\end{document}